\documentclass{ws-p9-75x6-50}

\begin{document}

\title{A Hybrid Cosmological Hydrodynamic/N-body Code Based on the Weighted
Essentially Non-Oscillatory Scheme}

\author{Long-Long Feng}

\address{Center for Astrophysics, University of Science
and Technology of China, Hefei 230026 \\ National Astronomical
Observatories, Chinese Academy of Science, Beijing, 100012}

\author{Chi-Wang Shu}

\address{Division of Applied Mathematics,
Brown University, Providence, RI 02912, U.S.A.}

\author{Mengping Zhang}

\address{Department of Mathematics,
University of Science and Technology of China, Hefei, Anhui
230026, P.R. China}

\maketitle

\abstracts{ We describe a newly developed cosmological
hydrodynamics code based on the weighted essentially non-oscillatory
(WENO) schemes for hyperbolic conservation laws.
High order finite difference WENO 
schemes are designed for problems with piecewise smooth
solutions containing discontinuities, and have been successful in
applications for problems involving both shocks and complicated
smooth solution structures. We couple hydrodynamics based on the
WENO scheme with standard Poisson solver - particle-mesh (PM)
algorithm for evolving the self-gravitating system.  A third
order total variation diminishing (TVD) Runge-Kutta scheme has
been used for time-integration of the system. We brief the
implementation of numerical technique. The cosmological
applications in simulating intergalactic medium and Ly$\alpha$
forest in the CDM scenario are also presented. }

\section{Introduction}

Due to the highly non-linearity of gravitational clustering in the
universe, one significant feature emerging in cosmological
hydrodynamic flow is extremely supersonic motion around the
density peaks developed by gravitational instability, which leads
to strong shock discontinuities within complex smooth structures.
It poses more challenge than the typical hydrodynamic simulation
without self-gravity. To address this issue, two algorithms have
been implemented for high resolution shock capturing in cosmology:
total-variation diminishing (TVD) scheme$^{[1]}$ and piecewise
parabolic method (PPM)$^{[2]}$.

An alternative hydrodynamic solver to 
discretize the convection terms in the Euler equations is the
fifth order finite difference WENO (weighted essentially
non-oscillatory) method$^{[3]}$, with a low storage
nonlinearly stable third order Runge-Kutta time discretization$^{[4]}$.
The WENO schemes, first constructed in third order finite volume
version by Liu et al.$^{[5]}$, are based on the essentially non-oscillatory
scheme (ENO) developed by Harten et al.$^{[6]}$ in the form of
finite volume scheme for hyperbolic conservative laws. The ENO
scheme generalizes the total variation diminishing (TVD) scheme of
Harten. The TVD schemes typically degenerate to first-order
accuracy at locations with smooth extrema while the ENO and WENO schemes
maintain high order accuracy there even in multi-dimensions.  
The WENO method is very robust and stable for
solutions containing strong shocks and complex solution
structures$^{[3],[5]}$. It uses the idea of adaptive stencils in the
reconstruction procedure based on the local smoothness of the
numerical solution to automatically achieve high order accuracy
and non-oscillatory property near discontinuities. This is
achieved by using a convex combination of a few candidate
stencils, each being assigned a nonlinear weight which depends on
the local smoothness of the numerical solution based on that
stencil.  WENO schemes can simultaneously provide a high order
resolution for the smooth part of the solution, and a sharp,
monotone shock or contact discontinuity transition.

In the context of cosmological applications, we have developed a hybrid
N-body/hydrodynamical code that incorporates a Lagrangian
particle-mesh algorithm to evolve the collision-less matter with
the fifth order WENO scheme to solve the equation of gas dynamics.
This paper describes this code briefly.

\section{Numerical Techniques}

\subsection{The Basic Equations}

The hydrodynamic equations for baryons in the expanding universe,
without any viscous and thermal conductivity terms, can be written
as,
\begin{equation}
\label{hydro} \dot{U}+\partial_i{\bf F}^i[U]=f(t,U)
\end{equation}
where the abbreviation $\partial_i\equiv\partial/\partial x^i$ has
been used, $x^i$ denote the proper coordinates. $U$ is a
five-component column vector, $U=(\rho,\rho v_1,\rho v_2, \rho
v_3, E)$, $\rho$ is the comoving density, ${\bf v}=\{v_i\}$ is the
proper peculiar velocity, E is the total energy per unit comoving
volume including both kinetic and internal energy, P is comoving
pressure, related to the total energy E by $
E=P/(\gamma-1)+\frac{1}{2}\rho {\bf v}^2$ where we assume an ideal
gas equation of state, $P=(\gamma-1)e$, where $e$ is the internal
energy density and $\gamma$ is the ratio of the specific heats of
the baryon. The left hand side of Eq.(1) is written in the
conservative form for mass, momentum and energy, the "force"
source term on right hand side includes the contributions from the
expansion of the universe, the gravitation. $\Lambda_{net}$ in the
energy equation represents the net energy loss due to the
radiative heating-cooling of the baryonic gas.
\begin{equation}
f(t,U)=(0,-\frac{\dot a}{a}\rho{\bf v}+\rho{\bf g}, -2\frac{\dot
a}{a}E+\rho{\bf v}\cdot{\bf g}-\Lambda_{cool})
\end{equation}
where ${\bf g}$ is the peculiar acceleration in the gravitational
field produced by both the dark matter and the baryonic matter,
which is obtained by solving the Poisson equation using standard
particle-mesh technique.

\subsection{Hydrodynamic Solver: Finite Difference WENO Schemes}

The fifth order WENO finite difference spatial discretization to a
conservation law such as
\begin{equation}
\label{1.1}
u_t + f(u)_x + g(u)_y = 0
\end{equation}
approximates the derivatives, for example
$f(u)_x$, by a conservative difference
$$
f(u)_x |_{x=x_j} \approx \frac{1}{\Delta x} \left(
\hat{f}_{j+1/2} - \hat{f}_{j-1/2} \right)
$$
where $\hat{f}_{j+1/2}$ is the numerical flux. $g(u)_y$ is
approximated in the same way.  Hence finite difference methods
have the same format for one and several space dimensions, which
is a major advantage.  For the simplest case of a scalar equation
(3) and if $f'(u) \geq 0$, the fifth order finite difference WENO
scheme has the flux given by
$$
\hat{f}_{j+1/2} = w_1 \hat{f}_{j+1/2}^{(1)}
+ w_2 \hat{f}_{j+1/2}^{(2)} + w_3 \hat{f}_{j+1/2}^{(3)}
$$
where $\hat{f}_{j+1/2}^{(i)}$ are three third order accurate fluxes on
three different stencils given by
\begin{eqnarray*}
\hat{f}_{j+1/2}^{(1)} & = &
\frac{1}{3} f(u_{j-2}) - \frac{7}{6} f(u_{j-1})
               + \frac{11}{6} f(u_{j}), \\
\hat{f}_{j+1/2}^{(2)} & = &
-\frac{1}{6} f(u_{j-1}) + \frac{5}{6} f(u_{j})
               + \frac{1}{3} f(u_{j+1}), \\
\hat{f}_{j+1/2}^{(3)} & = &
\frac{1}{3} f(u_{j}) + \frac{5}{6} f(u_{j+1})
               - \frac{1}{6} f(u_{j+2}).
\end{eqnarray*}
Notice that the combined stencil for the flux $\hat{f}_{j+1/2}$
is biased to the left, which is upwinding for the positive
wind direction due to the assumption $f'(u) \geq 0$.
The key ingredient for the success of WENO scheme relies on
the design of the nonlinear weights $w_i$, which are given by
$$
w_i = \frac {\tilde{w}_i}{\sum_{k=1}^3 \tilde{w}_k},\qquad
 \tilde{w}_k = \frac {\gamma_k}{(\varepsilon + \beta_k)^2} ,
$$
where the linear weights $\gamma_k$ are chosen to yield
fifth order accuracy when combining three third order
accurate fluxes, and are given by
$$
\gamma_1=\frac{1}{10}, \qquad \gamma_2=\frac{3}{5},
\qquad \gamma_3=\frac{3}{10} ;
$$
the smoothness indicators $\beta_k$ are given by
\begin{eqnarray*}
\beta_1 & = & \frac{13}{12} \left( f(u_{j-2}) - 2 f(u_{j-1})
                             + f(u_{j}) \right)^2 +
         \frac{1}{4} \left( f(u_{j-2}) - 4 f(u_{j-1})
                             + 3 f(u_{j}) \right)^2  \\
\beta_2 & = & \frac{13}{12} \left( f(u_{j-1}) - 2 f(u_{j})
                             + f(u_{j+1}) \right)^2 +
         \frac{1}{4} \left( f(u_{j-1})
                             -  f(u_{j+1}) \right)^2  \\
\beta_3 & = & \frac{13}{12} \left( f(u_{j}) - 2 f(u_{j+1})
                             + f(u_{j+2}) \right)^2 +
         \frac{1}{4} \left( 3 f(u_{j}) - 4 f(u_{j+1})
                             + f(u_{j+2}) \right)^2 ,
\end{eqnarray*}
and they measure how smooth the approximation based on a specific
stencil is in the target cell. Finally, $\varepsilon$ is a
parameter to avoid the denominator to become zero and is usually
taken as $\varepsilon = 10^{-6}$ in the computation.

This finishes the description of the fifth order finite difference
WENO scheme$^{[3]}$ in the simplest case.  As we can see, the
algorithm is actually quite simple and the user does not need to
tune any parameters in the scheme.

The WENO finite difference scheme is characterized by the
following properties. (1) The scheme is proven to be uniformly
fifth order accurate including at smooth extrema, and this is
verified numerically. (2) Near discontinuities the scheme produces
sharp and non-oscillatory discontinuity transition. (3) The
approximation is self-similar.  That is, when fully discretized
with the Runge-Kutta methods discussed in next section \S 2.3, the
scheme is invariant when the spatial and time variables are scaled
by the same factor. This is a major advantage for approximating
conservation laws which are invariant under such scaling. For the
details of how the scheme can be generalized in a more complex
situation, eventually to 3D systems such as Euler equations, we refer
to the reference [7].

\subsection{Time Discretizations}

To further discretize in time, we use a class of high order
nonlinearly stable Runge-Kutta time discretizations.  A
distinctive feature of this class of time discretizations is that
they are convex combinations of first order forward Euler steps,
hence they maintain strong stability properties in any semi-norm
(total variation norm, maximum norm, entropy condition, etc.) of
the forward Euler step, with a time step restriction proportional
to that for the forward Euler step to be stable, this proportion
coefficient being termed CFL coefficient of the high order
Runge-Kutta method$^{[8]}$. The most popular scheme in this class
is the following third order Runge-Kutta method for solving $u_t =
L(u,t)$, where $L(u,t)$ is a spatial discretization operator:
\begin{eqnarray}
\label{s1}
     u^{(1)} & = & u^n + \Delta t L(u^n, t^n)  \nonumber \\
     u^{(2)} & = & \frac{3}{4} u^n + \frac{1}{4} u^{(1)} + \frac{1}{4}
\Delta t L(u^{(1)}, t^n+\Delta t) \\
     u^{n+1} & = & \frac{1}{3} u^n + \frac{2}{3} u^{(2)} +
  \frac{2}{3} \Delta t L(u^{(2)}, t^n+ \frac{1}{2} \Delta t ) ,
\nonumber
\end{eqnarray}
which is nonlinearly stable with a CFL coefficient 1.0. However,
for our purpose of 3D calculations, storage is a paramount
consideration.  We thus use the third order low storage
nonlinearly stable Runge-Kutta method which was proven to be
nonlinearly stable in [4], with a CFL coefficient 0.32.

The timestep is chosen by the minimum value among three time
scales. The first is from the Courant condition given by
\begin{equation}
 \Delta t \le \frac{ CFL \times a(t) \Delta x}{\hbox{max}(|u_x|+c_s,
|u_y|+c_s, |u_z|+c_s)}
\end{equation}
where $\Delta x$ is the cell size, $c_s$ is the local sound speed,
$u_x$, $u_y$ and $u_z$ are the local fluid velocities and $CFL$ is
the Courant number, typically, we take $CFL=0.3$. The second
constraint is imposed by cosmic expansion which requires that
$\Delta a /a <0.02$ within single time step. The last constraint
comes from the requirement that a particle move no more than a
fixed fraction of cell size in one time step.

\section{Cosmological Application: IGM and Ly$\alpha$ Forest at High Redshift}

We run our hybrid N-body/hydrodynamic code to compute the cosmic
evolution of coupled system of both dark matter and baryonic
matter in a flat low density CDM model ($\Lambda$CDM), which is
specified by the cosmological parameters
$(\Omega_m,\Omega_{\Lambda},h,\sigma_8,\Omega_b)=(0.3,0.7,0.65,0.9,0.035)$.
The simulations were performed in a periodic, cubical box of size
12h$^{-1}$Mpc with a 128$^3$ grid and an equal number of dark
matter particles. Atomic processes including ionization, radiative
cooling and heating are modeled as in Cen$^{[9]}$ in a primeval
plasma of hydrogen and helium of composition ($X=0.76$, $Y=0.24$).
The uniform UV-background of ionizing photons is assumed to have a
power-law spectrum of the form $J(\nu) =J_{21}\times
10^{-21}(\nu/\nu_{HI})^{-\alpha}$ergs$^{-1}$cm$^{-2}$sr$^{-1}
$Hz$^{-1}$ parameterized by photoionizing flux $J_{21}$ at the
Lyman limit frequency $\nu_{HI}$, and is suddenly switched on at
$z\sim 6$ to heat the gas and reionize the universe.

\begin{figure}[t]
\begin{center}
\epsfxsize=30pc 
\epsfbox{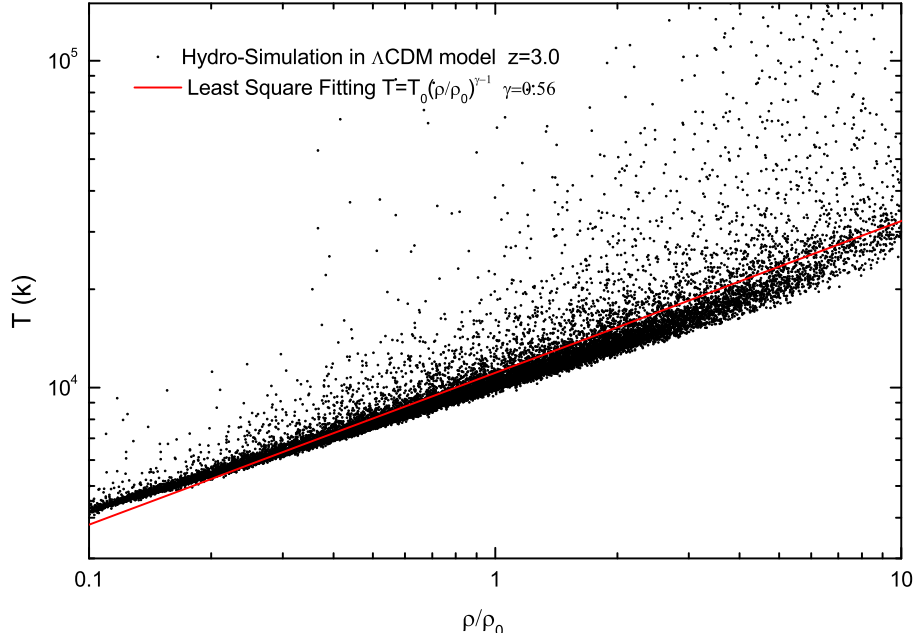} 
\caption{Scatter plot of IGM temperature vs. density relation
randomly drawn from a $\Lambda$CDM simulation at redshift $z=3.0$.
The line is the least-square-fitting of power-law using each set
of points displayed in the figure.}
\end{center}\end{figure}

The evolution of the IGM temperature relies on the reionization
history of hydrogen and helium, and plays important role in our
understanding the Ly$\alpha$ absorption spectrum. It is expected
that a power-law relation between IGM temperature and density
$T=T_0(\rho/<\rho>)^{\gamma-1}$ could be established due to the
ionizing photon heating the gas$^{[10]}$. We test T versus $\rho$
relation using the simulation samples.  Fig.(1) shows the scatter
plot of temperature-density relation drawn from the simulation at
the output of redshift z=3. Apparently, the T-$\rho$ relation
approximates roughly a power law with a best-fitting value of
$\gamma=0.56$. The dispersions in temperature at a fixed density
mainly arise from the shock-heating.

Mock Ly$\alpha$ spectra are produced along random lines of sight
in the simulation box, and meanwhile, the corresponding
one-dimension distributions of IGM density and peculiar velocity
are also extracted from the sample. The results are demonstrated
in Fig.(2). To compare with the observation, e.g. the Keck
spectrum of QSO q0014+8118, we Gaussian smooth the spectra to
match with the spectral resolutions of observation and normalize
the spectra by the observed mean flux decrement. For q0014+8118,
we have $D=1-<e^{-\tau}>\approx 0.31$. It could be shown that
there are good agreements of flux power spectrum and some
higher-order statistical behaviors$^{[11]}$ (e.g. intermittency)
between the observation and simulations$^{[12]}$.

\begin{figure}[t]
\begin{center}
\epsfxsize=27.2pc 
\epsfbox{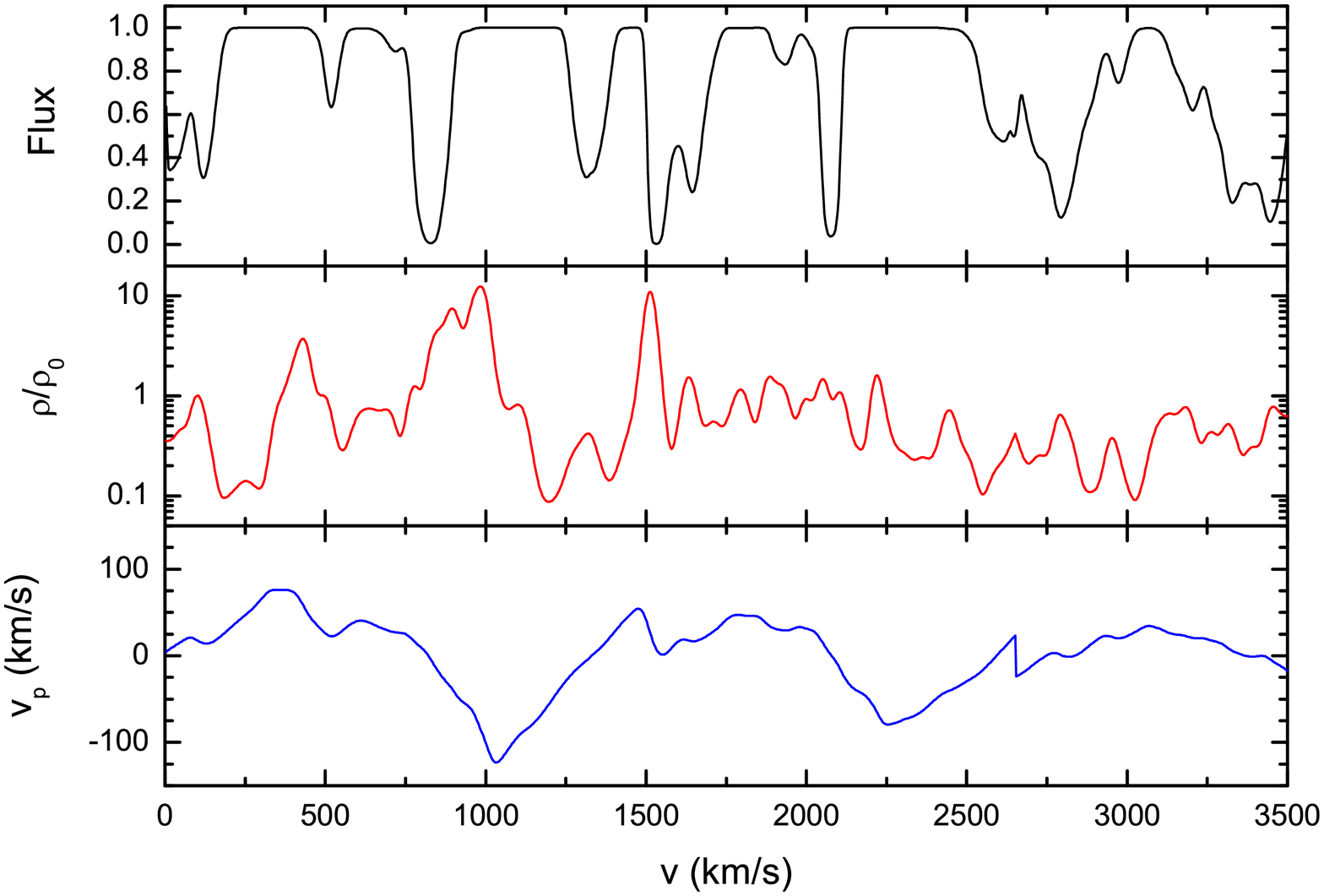} 
\caption{An example of simulated Ly$\alpha$ absorption spectrum and corresponding IGM physical
properties along randomly selected line of sight. Upper panel: transmitted flux; middle panel: density distribution
of the neutral hydrogen; lower panel: peculiar velocity distribution. \label{fig:lyaforest}}
\end{center}\end{figure}


\begin{thebibliography}{99}

\bibitem{}A. Harten, J. Comput. Phys., 49, 357 (1983)

\bibitem{}P. Collela \& P.R. Woodward, J. Comput. Phys., 54, 174 (1984)

\bibitem{}G. Jiang \& C.-W. Shu, J. Comput. Phys. 126, 202 (1996)

\bibitem{}S. Gottlieb \& C.-W. Shu, Math. Comp., 67, 73 (1998)

\bibitem{}X.-D. Liu, S. Osher \& T. Chan, J. Comput. Phys., 115, 200 (1994)

\bibitem{}A. Harten, B. Engquist, S. Osher \& S. Chakravarthy, J. Comput. Phys. 71, 231 (1987)

\bibitem{}C.-W. Shu,
in {\em Advanced Numerical Approximation of Nonlinear Hyperbolic
Equations}, B. Cockburn, C. Johnson, C.-W. Shu and E. Tadmor
(Editor: A. Quarteroni), Lecture Notes in
Mathematics,Springer,1697,325 (1998)

\bibitem{} S. Gottlieb, C.-W. Shu \& E. Tadmor, SIAM Review, 43, 89 (2001)

\bibitem{} R.Y., Cen, ApJS, 78, 341 (1992)

\bibitem{}L. Hui \& Gnedin, N.Y., MNRAS, 292,27 (1997)

\bibitem{} J. Pando,  L.L. Feng, P. Jamkhedkar, W. Zheng, D. Kirkman,
   D. Tytler  \& L.Z. Fang, ApJ, 574, 575 (2002)

\bibitem{} L.L. Feng, J. Pando \& L.Z. Fang, ApJ, accepted, (2002)

\end{thebibliography}
\end{document}